\documentclass[preprint]{emulateapj}

\slugcomment{Submitted to the Astronomical Journal}
\shortauthors{Eskew \& Zaritsky}
\shorttitle{}

\begin{document}
\title{NEARBY GALAXIES IN MORE DISTANT CONTEXTS}
  
\author{Michael Eskew and Dennis Zaritsky}
\affil{Steward Observatory, University of Arizona, 933 North Cherry Avenue, Tucson, AZ 85721, USA}

\email{meskew@email.arizona.edu, dzaritsky@as.arizona.edu}

\begin{abstract}    
We use published reconstructions of the star formation history (SFH) of the Large Magellanic Cloud (LMC), Small Magellanic Cloud, and NGC 300 from the analysis of resolved stellar populations  to investigate where such galaxies might land on
well-known extragalactic diagnostic plots over the galaxies' lifetime 
(assuming that nothing other than their
stellar populations change). For example, we find that the evolution of these
galaxies implies a complex evolution in the Tully-Fisher relation with lookback time and that 
the observed scatter is consistent with excursions these galaxies take as their stellar
populations evolve. 
We find that the growth of stellar mass is weighted to early times, despite the strongly 
star-forming current nature of the three systems. Lastly, 
we find that these galaxies can take circuitous paths across the color-magnitude diagram. 
For example, it is possible, within the constraints provided by the current determination of its SFH, that the LMC reached
the red sequence at intermediate age prior to 
ending back up on the blue cloud at the current time. Unfortunately, this behavior happens
at sufficiently early times that our resolved SFH is crude and insufficiently 
constraining to convincingly demonstrate that this was the actual evolutionary path. The limited
sample size precludes any general conclusions, but we present these as examples how we can bridge the study of resolved populations and the more distant universe.
\end{abstract}


\section{Introduction}
\label{sec:intro}

The universe has made it both easier and more difficult to study galaxy evolution than it 
could have. We are grateful for the finite
speed of light, which enables us to observe distant galaxies when they are young 
(in the cosmological sense).
On the other hand, we are frustrated (at least in this sense) with the grave mismatch in galactic and human timescales that precludes us from observing any individual galaxy as it evolves. 
As such, we have 
relied on observing changes in galaxy ensembles \citep[cf.][]{bo} across cosmic time 
and testing simple
evolutionary models against such observations \citep[cf.][]{tinsley}. These theoretical exercises are
statistical consistency checks 
and do not directly verify the unique ability of any particular model to represent galaxy evolution. 

How convoluted could the actual evolutionary history of each individual galaxy be? To study the 
evolution of a single galaxy requires us at a minimum to understand the rate at which it formed stars over
its history. Of course, there are other aspects of a galaxy's evolution, including the accretion of both
dark matter and baryons, that are harder to constrain. Even so, the relatively straightforward study
of the star formation history (SFH) of galaxies has been fairly limited to date. Due to technical limitations, such work
has focused almost exclusively on Local Group galaxies \citep{tolstoy}, although the superior angular resolution of
space-based telescopes has allowed us to begin a systematic extension to galaxies outside the Local Group \citep[for example, see][]{dalcanton}. This extension is critical
because the galaxies one can study at larger distances are intrinsically brighter (by many magnitudes) than
the typical Local Group galaxy, and therefore more representative of unresolved galaxies studied at even
larger distances. Bridging the study of a relatively small number of resolved galaxies and that of statistically large samples at higher redshift has proved difficult.  Here 
we examine how a small set of local galaxies behave as their stellar populations evolve on some 
well-known diagnostic plots that are often used to characterize unresolved galaxies.
The aim of this exercise is to begin formulating intuition about
the nature of the scatter in these relations and any
systematic behavior with time.
We focus on the two 
most luminous Local Group galaxies for which global SFHs have been 
measured \citep[the Large Magellanic Cloud (LMC) and the Small Magellanic Cloud (SMC);][]{hz04,hz09} and one additional
relatively massive, nearby galaxy for which a nearly global history has been published \citep[NGC 300;][]{gogarten}. 

\section{Reconstructed Galaxy Evolution}

We begin with the published star formation rates (SFRs) as a function of  time 
for the three galaxies of interest.
These are then converted into luminosities, colors, and stellar masses over time using
the population synthesis code, P\'EGASE
\citep[][although we use their updated Version 2 from 2001]{pegase}.
P\'EGASE produces a historical record of observable galactic characteristics using the 
input galactic SFH and specified input parameters, 
such as the stellar initial mass function. Unless otherwise stated, we adopt the default parameter values. Given the limited number of galaxies in our sample, this is not a parameter study and we opt for simplicity.
There are, of course, various limitations that will eventually restrict what one can do with this
approach. The most important of these is that the temporal resolution of the SFHs is itself a function of time and generally degrades, resulting in 
temporal resolution  that is roughly constant with log(age).

For most general studies of galaxy evolution that use P\'EGASE, the adopted input SFH is specified using a simple analytical
model such as an exponentially declining SFR (the $\tau$ models).
Fortunately, for generality, the algorithm was constructed to allow for non-parametric,
user-defined histories. We use the source SFHs as published for the LMC \citep{hz09}, SMC \citep{hz04}, and NGC 300 \citep{gogarten}. For the LMC and SMC we combine the SFHs from all spatial bins
to produce a single, global SFH. For NGC 300 we also combine SFHs from the spatial bins, but because the coverage is not complete we normalize each
by the inverse of the fractional area observed at that radius to obtain an estimated global SFH. 
This ratio is calculated using the size of the observed field and the area of the full aperture
at the given radius.
We ignore the metallicity information provided by these sources (thereby allowing P\'EGASE to
calculate an internally consistent chemical evolutionary history).  The assumption of
internally self-consistent evolution has been shown to fail in detail for the SMC by \cite{zh}, which
they interpret as the need for gaseous infall, but we ignore that additional level of complication
here.

In detail, some slight adjustments to the published SFHs are needed in the interest of 
uniformity.
We modify each SFH to cover 14 Gyr by extending or contracting the initial time bin in the published versions when necessary, proportionately decreasing or increasing the SFR so that the total number of stars produced remains unchanged. 
There remains a slight tension between ages determined from stellar models and those from the currently standard Lambda cosmology ($\Omega_m = 0.27$, $\Lambda = 0.73, H_0 = 70$ km s$^{-1}$ Mpc$^{-1}$) in that the oldest isochrones used tend to be older than the age of the universe. However, a modest change in the value of $H_0$ to 67, for example, allows ages up to 14.5 Gyr, which we use as our upper limit. Any stars formed from isochrones with stated ages older than 14.5 Gyr are incorporated
into this oldest bin. Our temporal resolution at these ages is extremely coarse, so we are insensitive to 
such distinctions although they remain a numerical oddity. Alternatively, we also ran models where every age bin was narrowed by a constant fraction to provide consistency between 
the total age and 
cosmological constraints. These models produced no significant differences, with one
exception that we discuss further in Section 3.2. In all other cases, we present the results only from the first approach.

To produce a relatively accurate self-consistent model, the algorithm must account properly
for both the stars formed and the gas consumed (and recycled). The model naturally begins with
all of its mass in the gas phase, but the gas mass relative to the final stellar mass is a free
parameter. A given SFH could result either in a gas-rich or gas-poor galaxy depending on how
much gas the system begins with. To remove
this degeneracy, we require that the model produce the current observed gaseous to stellar mass ratio for each galaxy. The P\'EGASE models work on the basis of evolving 1 M$_\odot$ of gas. We scale
the SFH so that the final ratio of mass in gas to stars matches observations. We adopt measurements of the current 
gas masses from \cite{stavely} and \cite{mizuno} for the LMC, \cite{tumlinson} for the SMC, and \cite{deV}
for NGC 300.
Lastly, we scale
the mass of the entire system so that the final stellar mass matches the integral of the published
SFHs.

Once these quantities are determined, P\'EGASE straightforwardly calculates the colors,
luminosities, and stellar masses at each age. These output values enable us to place
our galaxies on the diagnostic plots.

\section{Discussion}

With the stellar population evolutionary histories of these galaxies in hand, we now investigate
how these galaxies move across three different, but common, diagnostic plots. 
First, we explore their behavior
on the Tully-Fisher relationship \citep{tf}. This scaling law is commonly used to explore disk galaxy evolution,
with the expectation that systemic evolution of the sizes or luminosities of the galaxies might lead
to zero point and/or slope changes in the relationship \citep[cf.][and references therein]{weiner}. Second, we explore their
behavior on the stellar-mass density versus redshift (or age) plot. Ensembles of galaxies across 
many redshifts are used to construct the plot that shows when the universe creates it stars \citep[cf.][and references therein]{rudnick06}. Of course, such measurements represent ensemble averages, so it is worthwhile to compare with how individual galaxies build their stellar populations. 
This exercise is being done for a large sample of dwarf galaxies \citep{weisz} and our
sample overlaps that one slightly and extends the coverage toward more massive systems. 
In particular, there are claims of generic galaxy behavior in which low-mass galaxies form
more of their stars later (more recently) than more massive galaxies \citep[``downsizing";][]{cowie}. 
However, the term ``downsizing"  has confusingly 
been used in many different ways \cite[for an explanation of various forms of downsizing see][]{fontanot}. 
We will test whether such trends are evident for at least these three galaxies. Lastly, we explore the galaxies' behavior
on the color-magnitude diagram of galaxies. Here the questions relate primarily to how or 
whether galaxies migrate between the various identified populations (eg. red sequence, green valley, blue cloud). The general expectation is that galaxies migrate (eventually) from the blue cloud
through the green valley onto the red sequence, although all three of our galaxies are currently 
star forming.

\subsection{The Tully-Fisher Relation}

Using the results from the P\'EGASE modeling, we present the evolution of the three galaxies
on the TFR and compare them with the current positions of 
162 spiral galaxies \citep{pizagno} and 
of 87 isolated, low-mass galaxies \citep{blanton}.
Rotation values for the LMC \citep{alves}, SMC \citep{stan}, and NGC 300 \citep{puche} are taken to be constant in time. 

\begin{figure}[htbp]
\begin{center}
\plotone{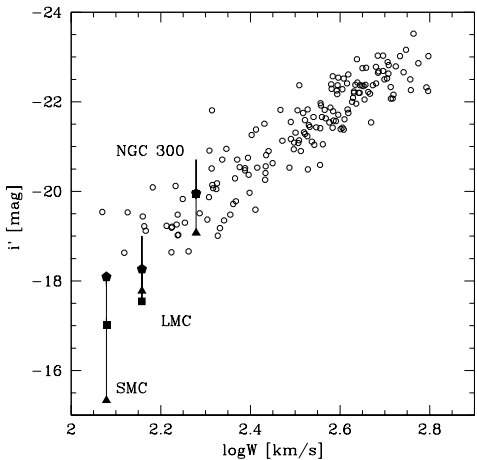}
\caption{Evolution of galaxies on the Tully-Fisher relation. Open symbols represent the 162 SDSS galaxies studied in the compilation by \cite{pizagno}. The solid symbols represent the SMC, LMC, and NGC 300 with the solid line tracing their evolution and the symbols representing different ages (triangle: 0.5 Gyr; square: 8 Gyr; pentagon: current-day).}
\label{fig:tfr}
\end{center}
\end{figure} 

The current positions of the three galaxies fall well within the scatter extrapolated from the
distribution of spiral galaxies (Figure \ref{fig:tfr}), suggesting that there is no fundamental problem with the calculated current-day magnitudes from our modeling. 
The SMC lies at the lower end of the range in rotation speeds for the Sloan Digital Sky Survey (SDSS) galaxies, highlighting how poorly most Local Group galaxies match those studied beyond the Local Group. 
Even
our most massive galaxy, NGC 300, does not reach the median rotation value of this local sample. The problem
becomes increasingly acute when studying cosmologically interesting samples.

The LMC and NGC 300, the two galaxies that are somewhat more representative of
the SDSS sample, spend their entire lifetime within the scatter of the current-day positions of the SDSS galaxies. Assuming
that the fluctuations in star formation experienced by these galaxies is not synchronized, 
this finding suggests that much of the observed TFR scatter could be due to variations in 
stellar populations. In contrast, investigators have searched for correlations between TFR residuals and
color \citep{kannappan,pizagno} and found little or none. It is unclear the extent to which
that type of investigation is complicated by internal extinction and geometry, and to which
our results may not be representative of more massive (typical) TFR galaxies.

\begin{figure}[htbp]
\begin{center}
\plotone{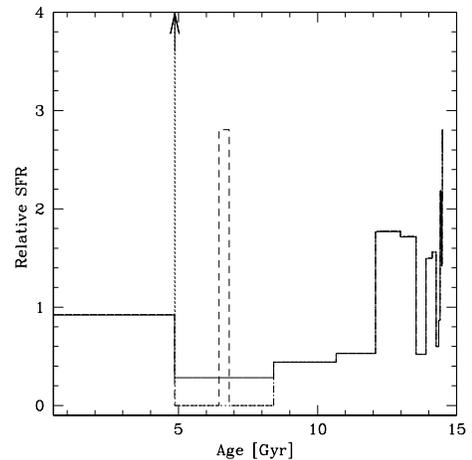}
\caption{Star formation history of the LMC. The solid line represents the baseline model
of the star formation history of the LMC (see \cite{hz09} for details). Current-day is at the right of the figure.
The dotted line represents Burst Model 1 (BM1, see the text for details) with a burst at $\sim$ 5 Gyr. This burst reaches $\sim$ 66 on the y-axis. The dashed line represents Burst Model 2 (BM2) with a longer period of more modestly elevated star formation at $\sim$ 6.5 Gyr.}
\label{fig:bursts}
\end{center}
\end{figure}

Although the SMC shows a systematic brightening over time, as one might expect as
a galaxy progressively forms more stars, both the LMC and NGC 300 are not currently
at their brightest luminosity. 
These two galaxies move up and down within the observed TFR scatter
over their lifetime, suggesting that there may not be a 
simple coherent evolution of the TFR with redshift.  On the other hand, even if
the evolution is not monotonic, the TFR may still be a powerful diagnostic to the degree that evolution is synchronized for galaxies of 
a given mass. For example, in a scenario where the massive galaxies form most of their stars first and the intermediate galaxies, such as NGC 300 and the LMC, are more vigorously forming stars now, one might expect the TFR to be steeper in the 
distant past relative to the current TFR. Interestingly in this
regard, the luminosities of the LMC and NGC 300 do behave similarly with time.

One important hidden aspect of the SFHs we use is that we
only constrain the number of stars formed over a fairly large amount of time, particularly for
ages  $>$ few Gyr. Within a bin, we have no constraint on the fluctuations
in the SFR. 
To study possible effects of  unresolved bursts, we create two models for the LMC that probe a range of 
possibilities. 
In our  Burst Model 1 (BM1), we place all of the star formation
in our second oldest bin, which happens to be the one with the longest age span, into as small a burst as we ever resolve ($\sim$ 20 Myr). 
This is evidently a completely unrealistic, extreme scenario, but it illustrates the maximum impact that
an unresolved, ancient burst would have on our results.  In our Burst Model 2 (BM2) we
create a modestly elevated SFR, corresponding in magnitude to what is typical at more recent times
where we have better resolution. We define the length of that ``burst" so that it produces the
same total number of stars over the corresponding bin as in our baseline SFH. 
This scenario represents the smallest unresolved 
variation that we should expect.
These two burst models are graphically
illustrated in Figure \ref{fig:bursts}.

\begin{figure}[htbp]
\begin{center}
\plotone{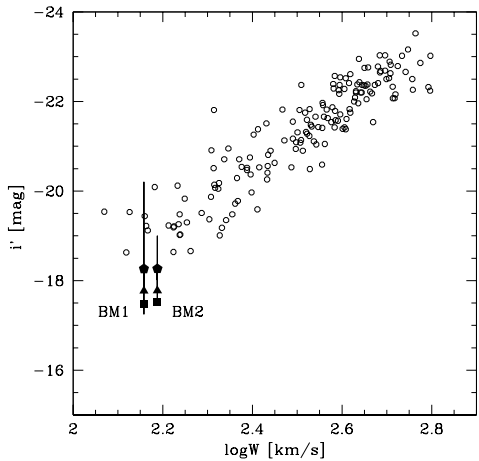}
\caption{Effect of star formation bursts on the Tully-Fisher relation. Symbols represent the same as in 
Figure \ref{fig:tfr} although the LMC evolution is now shown for the Burst Model 1 (BM1; left) and Burst Model 2
(BM2; right). BM1 represents an extreme scenario that is unrealistic but is meant to explore
the maximum effect of unresolved star formation episodes. BM2 is a more realistic model meant to match the variation in star formation rates seen currently. The two models are displaced horizontally for clarity, with BM1 being at the correct location for the LMC.}
\label{fig:tfburst}
\end{center}
\end{figure}

The results of these burst models are that the TFR evolution is likely to be qualitatively unaffected by
unresolved bursts.
BM2 produces a TFR history (Figure \ref{fig:tfburst}) that is effectively indistinguishable from that shown in Figure \ref{fig:tfr}. 
Even for  BM1, the LMC's luminosity, while noticeably larger than in the baseline model,  still places the galaxy within 
the TFR scatter. If the TFR scatter has several contributing components, then it
will be difficult to use residuals from the TFR to infer variations in SFR of a scale
seen in the LMC and NGC 300. The apparent ``scatter" due to SFH variations will be present regardless of what additional complications, 
such as mass accretion, are occurring.

\begin{figure}[htbp]
\begin{center}
\plotone{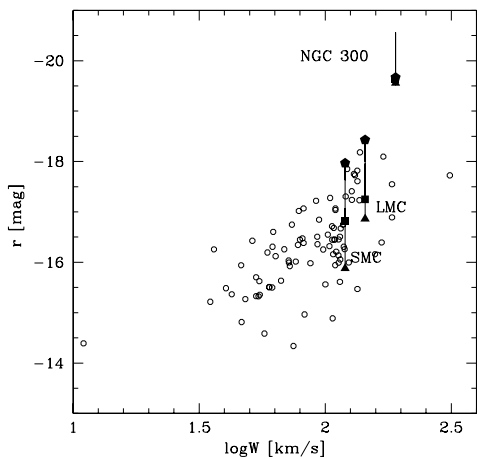}
\caption{Evolution of galaxies on the Tully-Fisher relation. Open symbols represent the 87 galaxies studied in the compilation by \cite{blanton}, corrected to match the gas fraction of
the Clouds (see the text for details). The solid symbols represent the SMC, LMC, and NGC 300 with the solid line tracing their evolution and the symbols representing different ages {(triangle: 0.5 Gyr; square: 8 Gyr; pentagon: current-day).}
}
\label{fig:new-tf}
\end{center}
\end{figure}

One shortcoming of the previous comparison is the limited overlap between the \cite{pizagno}
galaxies and our sample of three, particularly with the two Magellanic Clouds. In an effort
to address this issue, we utilize the data from \cite{blanton}, who investigated the TFR
for low-luminosity, isolated galaxies. By its nature, this sample consists of fairly gas-rich galaxies
that often have the majority of their baryons in the gas phase rather than the stellar phase. This
characteristic complicates the comparison to our sample because the Clouds are relatively stellar dominated (both have roughly 75\% of their baryons in the stellar phase). To mitigate this difference, we ``convert" a fraction of the gas in the \cite{blanton} galaxies into stars, using the $M/L$ value adopted by \cite{blanton}, and by doing so calculate the luminosity of these galaxies if they too had 75\% of their baryons in the stellar phase (this is analogous to placing galaxies on a baryonic TF; see \cite{btf}). The comparison of their TFR to our galaxies is shown in Figure \ref{fig:new-tf}. This comparison leads to the same conclusions as those described previously 
for the comparison to the \cite{pizagno} galaxies, although here our galaxies are at the upper end of the
$\log(W$) range.

\subsection{Stellar Mass Evolution}

\begin{figure}[htbp]
\begin{center}
\plotone{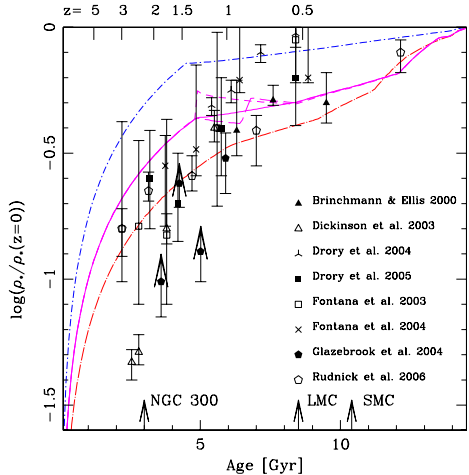}
\caption{Growth of stellar mass. Symbols represent data from ensembles of high-redshift
galaxies taken from the cited studies and the format of the plot follows \cite{rudnick06}. The lines represent the results for our three galaxies: the LMC, NGC 300, and SMC going from highest
to lowest at $z = 5$. Additionally, the line for each galaxy is identified by color in the online version of the plot: LMC tracks are in magenta, SMC track in red, and NGC 300 track in blue. The two dashed lines tracing the baseline LMC result represent two different alternate SFHs for the LMC (BM1 and BM2, see the text for details). The labeled arrows at the bottom mark the time by which 50\% of the current stellar mass has formed.}
\label{fig:rho}
\end{center}
\end{figure}

The opening of the higher redshift ($z > 0.5$) universe of galaxies to observations 
has provided clear evidence
for the modulation in the overall rate of star formation with lookback time \citep{lilly, madau}.
Common diagnostic plots of the stellar populations in galaxies now include 
both the differential (the SFR or ``Madau" plot) and 
the integral form (the stellar mass density). 
In Figure \ref{fig:rho}, we plot the fraction of the stellar mass formed as a function of time for each galaxy, including the baseline LMC model and the two burst models, and compare to several sets of published data for the stellar mass density as measured from ensembles of high-redshift galaxies. This comparison, specifically for the LMC, is the only case where our treatment for
addressing the internal inconsistency between our oldest stellar populations and the age 
of the universe results in significantly different results. Using the approach where we simply truncate
the oldest bin results in the LMC being discrepant with the high-redshift sample
at a level comparable to that for NGC 300, whereas the curves shown in Figure \ref{fig:rho} correspond to the approach where we proportionally trim the length
of each time bin. The alternative approaches do not result in significant differences for NGC 300 and the SMC.

It is evident that NGC 300 forms fractionally more stars at early times than the bulk of the cosmologically measured galaxies. The SMC and LMC either follow the high-redshift trend or lie only slightly below it. The sense of these results runs counter to a simple interpretation of the ``downsizing" concept because these relatively low-mass galaxies did not form their stars at more recent times
than the bulk of higher mass galaxies observed at high redshifts. However, 
our sample of 
three does indeed follow the trend in that the stellar fraction is always higher in the more massive galaxy.

We conclude that we see few or no signs from this small sample that low-mass galaxies have 
necessarily formed a larger fraction of their stellar populations more recently than did higher mass
galaxies, unless one refers to SMC-like and lower mass galaxies as those participating in the 
phenomenon. Of course, the LMC and SMC are perhaps unusual in that they are
close to the Milky Way, although some suggest that the LMC/SMC are on
their first approach toward the Milky Way \citep{besla}. There are ways around the apparent
paradox. For example, if the ensemble work is missing a significant ancient stellar population in
galaxies, then it has an incorrect normalization relative to our galaxies. 
Missing such a population
would not necessarily affect the relative comparison of massive and less massive galaxies within
the ensembles, but would distort the comparison to our galaxies. 
         
\subsection{The Color-Magnitude Diagram}

The color-magnitude diagram exemplifies the interesting bimodality of galaxies. The red
sequence contains the non-star-forming, generally pressure-supported galaxies, while
the blue cloud contains the star-forming, generally rotationally supported galaxies. Although
some exceptions exist \citep{chen}, the general evolution model that is envisioned \citep[cf.][]{bell}
is that 
at some point in their history, perhaps as a result of an interaction, blue cloud galaxies
move toward the red sequence passing through the intermediate region termed the
green valley. This is a fairly orderly picture of galaxy evolution and is supported by the
observed growth in the number of galaxies on the red sequence, particularly at fainter
magnitude \citep{delucia}. However, individual galaxies may not be so well behaved.

\begin{figure}[htbp]
\begin{center}
\plotone{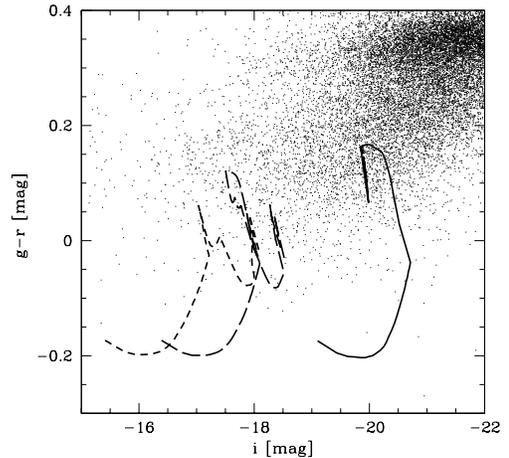}
\caption{Evolution of galaxies on the color-magnitude diagram. We plot the evolutionary
paths of the SMC, LMC, and NGC 300 from left to right, respectively, on the distribution of local galaxies from
SDSS. All galaxies begin at the lower left extreme of their track and move as time progresses toward the blue cloud. The evolutionary paths are mostly constrained to lie in the ``blue cloud" of galaxies and all galaxies are currently in the blue cloud.}
\label{fig:cmd}
\end{center}
\end{figure}

In Figure \ref{fig:cmd}, we show the color and magnitude evolution of 
the LMC, the SMC, and NGC 300 in comparison to the current-day distribution of over $4\times 10^5$ SDSS galaxies \citep{blanton}. The evolutionary
tracks show that after the initial phase, which is a blue hook and not well constrained
by the available SFHs, the galaxies all progress onto the blue cloud. Their evolutionary
paths take them through the large range of colors seen in the blue cloud, but not beyond
it. As such, we might conclude that the picture of galaxies remaining in the blue cloud
prior to any possible, eventual migration onto the red sequence is correct. However,
these models represent the minimum excursion set because we average the SFR within a given
time bin. 

\begin{figure}[htbp]
\begin{center}
\plotone{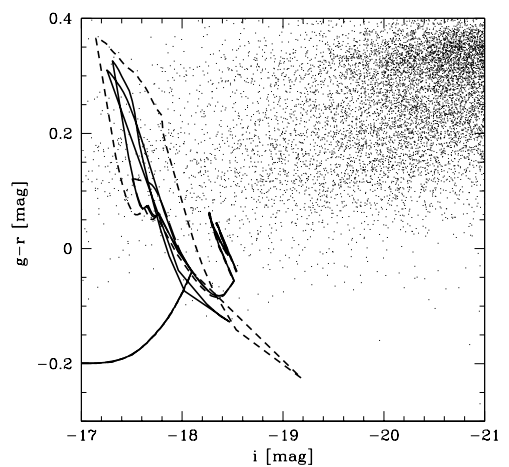}
\caption{Different possible evolutionary paths for the LMC 
on the galaxy color-magnitude diagram. We plot the evolutionary
paths of the LMC in the baseline model (long-dashed line), BM1 (short-dashed line),
and BM2 (solid line). Both alternate models have the LMC reaching colors as red as
the red sequence before returning to the blue cloud once star formation starts again.
}
\label{fig:cmdlmc}
\end{center}
\end{figure}

We show a range of possible behavior for the LMC in Figure \ref{fig:cmdlmc}. We plot our baseline
model, BM1, and BM2. 
BM1 with its rather extreme burst
shows a path by which the LMC would reach the red sequence prematurely. This result may at first seem contradictory in that adding a burst has enabled the LMC to reach the red sequence. This behavior occurs because we have a constraint on the total SFR within the bin, and therefore, placing a burst at one time implies a deficit in star formation at all other times.
Of course, BM1 is an unrealistic, extreme scenario, but 
BM2, which is consistent in amplitude with the bursts seen at recent times, also takes the LMC to the red sequence. 

We conclude that our SFHs do not exclude eras of quiescence that would allow a galaxy like
the LMC to evolve from the blue cloud onto the red sequence, only to have it later evolve back into the blue cloud once star formation sufficiently 
re-engages. This exercise highlights both the potential and limitations of the approach we
are taking. At face value, the current SFHs support a model in which these galaxies have
always been among the blue cloud galaxies. However, the current data are insufficient to answer the question of whether these galaxies
wandered onto the red sequence only to return to the blue cloud. Deeper, higher resolution
data from which higher temporal resolution SFHs could be constructed would
help address this question. Nevertheless, it is evident that galaxies will move significantly, and 
stochastically, as they evolve within this diagram.

\section{Summary}

We illustrate how SFHs, reconstructed from the resolved stellar populations of
nearby galaxies, can be used to build intuition regarding the behavior of individual galaxies in
diagnostic plots used more commonly to study the behavior of galaxy ensembles at greater
distances. Specifically, we investigate the behavior of the SMC, LMC, and
NGC 300 in the TFR, the stellar mass versus cosmic time plot, and the color-magnitude diagram of galaxies.

The three galaxies we investigate all currently fall within the observed local TFR. However, their
evolution drives them across the full range of the scatter, suggesting that variation in the SFH contributes
significantly to the observed scatter. We also find that the evolution we measure in the 
stellar populations does not result in a
monotonic drift relative to the relation, thereby suggesting that an ensemble of galaxies may not
show evident zero point or slope changes over redshift, or that those changes may be weaker than
otherwise expected. 

The three galaxies we investigate all show stronger early growth in their stellar mass than the
average for field galaxies. This result is also found for a larger sample of lower mass, local dwarf galaxies by
\cite{weisz}. This result brings to question a simple model of ``downsizing" where low-mass galaxies
form a larger fraction of their stars after the more massive galaxies. While the current SFRs of the
three galaxies here place them all among blue, star-forming galaxies, the data show that 
star formation in these systems was not wholly delayed until relatively recent times.

The three galaxies we investigate all trace paths that populate the blue cloud of the color-magnitude
diagram. However, at least one plausible SFH, in that it is allowed within the range of uncertainties in the current measurement of the LMC SFH, has the LMC reaching the red sequence before returning to its current location among
the blue cloud galaxies. As such, we cannot exclude the possibility that galaxies wander across
populations in the diagram rather than systematically proceeding from blue cloud to red sequence.

The current study is limited by sample size, temporal resolution, and the lack of
higher mass galaxies that are more representative of the class of galaxy observed at higher redshifts.
Nevertheless, it exemplifies some of what might be possible when resolved SFHs are used to examine the behavior of individual galaxies on common diagnostic plots.

\begin{acknowledgments}

We thank S. Gogarten for communicating her star formation rate measurements for
NGC 300 electronically and acknowledge financial support from NASA LTSA  award 
NNG05GE82G and NSF grants AST-0307482 and AST-0907771.

\end{acknowledgments}

\end{document}